%% file: model.tex
\documentclass[onecolumn, journal]{IEEEtran}
 
\input{Preamble/preamble}

\begin{document}

\title{\LARGE{Synthetic 33-Bus Microgrid: Dynamic Model and Time-Series Parameters}}

\author{Tong Han
 
\vspace{-20pt}
}

\maketitle

\section{Introduction}

This report provides the detailed description of the synthetic 33-bus microgrid (MG), including its structure, dynamic models, and time-series parameters of loads and generations. The network structure is adapted from the IEEE 33-bus distribution network, with additional converter-interfaced renewable energy resources and energy storage systems. Time-series parameters is generated based on the open-source ARPA-E PERFORM datasets \cite{4-1668}.

\section{Basic structure}

\begin{figure}[h]
	\centering
 	\includegraphics[scale=1.2]{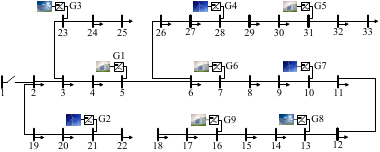}  
	\caption{Diagram of the synthetic 33-bus microgrid.}
	\label{fig-7-1-2}
\end{figure}

The single-line diagram of the 33-bus MG is shown in Fig. \ref{fig-7-1-2}.  
The entire MG contains 23 loads, and 9 converter-interfaced generations, named G1 to G9, including 3 wind generators, 2 solar panel generators, and 4 energy storage systems. 
 
\section{Dynamic Model with hierarchical control and renewable uncertainties}

\subsection{Converter model}

The dynamics of the converters with hierarchical control, also including renewable uncertainties, can be formulated as below:
\begin{subequations}\label{eq-7-1-1}
    \begin{align}
        & \bm{\omega} = \omega^* - \bm{K}_{\rm p}( \bm{P} - (\bm{P}^* + \bm{\xi})) + \bm{u}_{\rm p} - \omega_{\rm ref}  \\
        & \sqrt{(\bm{v}_{\rm od}^*)^2 + (\bm{v}_{\rm oq}^*)^2 } = \bm{U}^* - \bm{K}_{\rm q} (\bm{Q} - \bm{Q}^*) + \bm{u}_{\rm q} \\
        & \dot{\bm{\delta}} = \bm{\omega} - \omega_{\rm n} \\
        & \dot{\bm{P}} = \omega_{\rm c} ( - \bm{P} + \bm{v}_{\rm od} \sodot \bm{i}_{\rm od} + \bm{v}_{\rm oq} \sodot \bm{i}_{\rm oq}) \\
        & \dot{\bm{Q}} = \omega_{\rm c} ( - \bm{Q} + \bm{v}_{\rm oq} \sodot \bm{i}_{\rm od} - \bm{v}_{\rm od} \sodot \bm{i}_{\rm oq}) \\
        & \dot{\bm{\phi}_{\rm d}} = \bm{v}_{\rm od}^* - \bm{v}_{\rm od}  \\
        & \dot{\bm{\phi}_{\rm q}} = \bm{v}_{\rm oq}^* - \bm{v}_{\rm oq}  \\ 
        & \bm{i}_{\rm ld}^* = \bm{F} \sodot \bm{i}_{\rm od} \!-\! \omega_{\rm n} \bm{C}_{\rm f}  \sodot \bm{v}_{\rm oq} \!+\! \bm{K}_{\rm pv} \sodot (\bm{v}_{\rm od}^* \!-\! \bm{v}_{\rm od}) \!+\! \bm{K}_{\rm iv} \sodot \bm{\phi}_{\rm d} \\
        & \bm{i}_{\rm lq}^* = \bm{F} \sodot \bm{i}_{\rm oq} \!+\! \omega_{\rm n} \bm{C}_{\rm f}  \sodot \bm{v}_{\rm od} \!+\! \bm{K}_{\rm pv} \sodot (\bm{v}_{\rm oq}^* \!-\! \bm{v}_{\rm oq}) \!+\! \bm{K}_{\rm iv} \sodot \bm{\phi}_{\rm q} \\
        & \dot{\bm{\gamma}_{\rm d}} = \bm{i}_{\rm ld}^* - \bm{i}_{\rm ld}  \\
        & \dot{\bm{\gamma}_{\rm q}} = \bm{i}_{\rm lq}^* - \bm{i}_{\rm lq}  \\ 
        & \bm{v}_{\rm id}^* = - \omega_{\rm n} \bm{L}_{\rm f} \sodot \bm{i}_{\rm lq} + \bm{K}_{\rm pc} \sodot (\bm{i}_{\rm ld}^* - \bm{i}_{\rm ld}) + \bm{K}_{\rm ic} \sodot \bm{\gamma}_{\rm d} \\
        & \bm{v}_{\rm iq}^* = \omega_{\rm n} \bm{L}_{\rm f} \sodot \bm{i}_{\rm ld} + \bm{K}_{\rm pc} \sodot (\bm{i}_{\rm lq}^* - \bm{i}_{\rm lq}) + \bm{K}_{\rm ic} \sodot \bm{\gamma}_{\rm q} \\
        & \dot{\bm{i}_{\rm ld}} = - \bm{r}_{\rm f} \soslash \bm{L}_{\rm f} \sodot \bm{i}_{\rm ld} + \omega_{\rm n} \bm{i}_{\rm lq} + \bm{v}_{\rm id}^* \soslash \bm{L}_{\rm f} - \bm{v}_{\rm od} \soslash \bm{L}_{\rm f} \\
        & \dot{\bm{i}_{\rm lq}} = - \bm{r}_{\rm f} \soslash \bm{L}_{\rm f} \sodot \bm{i}_{\rm lq} - \omega_{\rm n} \bm{i}_{\rm ld} + \bm{v}_{\rm iq}^* \soslash \bm{L}_{\rm f} - \bm{v}_{\rm oq} \soslash \bm{L}_{\rm f}  \\
        & \dot{\bm{v}_{\rm od}} = \omega_{\rm n} \bm{v}_{\rm oq} + \bm{i}_{\rm ld} \soslash \bm{C}_{\rm f} - \bm{i}_{\rm od} \soslash \bm{C}_{\rm f} \\
        & \dot{\bm{v}_{\rm oq}} = - \omega_{\rm n} \bm{v}_{\rm od} + \bm{i}_{\rm lq} \soslash \bm{C}_{\rm f} - \bm{i}_{\rm oq} \soslash \bm{C}_{\rm f} \\
        & \dot{\bm{i}_{\rm od}} = - \bm{r}_{\rm c} \soslash \bm{L}_{\rm c} \sodot \bm{i}_{\rm od} + \omega_{\rm n} \bm{i}_{\rm oq} + \bm{v}_{\rm od} \soslash \bm{L}_{\rm c} - \bm{E}_{\rm C} \bm{v}_{\rm Bd} \soslash \bm{L}_{\rm c} \\
        & \dot{\bm{i}_{\rm oq}} = - \bm{r}_{\rm c} \soslash \bm{L}_{\rm c} \sodot \bm{i}_{\rm oq} - \omega_{\rm n} \bm{i}_{\rm od} + \bm{v}_{\rm oq} \soslash \bm{L}_{\rm c} - \bm{E}_{\rm C} \bm{v}_{\rm Bq} \soslash \bm{L}_{\rm c}
    \end{align}
\end{subequations}

Eliminating the algebraic variables associated with converters gives
{
\small{
\begin{subequations}\label{eq-7-1-2}
    \begin{align} 
        & \dot{\bm{\delta}} = \omega^* - \bm{K}_{\rm p}( \bm{P} - (\bm{P}^* + \bm{\xi})) + \bm{u}_{\rm p} - \omega_{\rm ref}  \label{eq-7-1-2:1}\\
        & \dot{\bm{P}} = \omega_{\rm c} ( - \bm{P} + \bm{v}_{\rm od} \sodot \bm{i}_{\rm od} + \bm{v}_{\rm oq} \sodot \bm{i}_{\rm oq}) \\
        & \dot{\bm{Q}} = \omega_{\rm c} ( - \bm{Q} + \bm{v}_{\rm oq} \sodot \bm{i}_{\rm od} - \bm{v}_{\rm od} \sodot \bm{i}_{\rm oq}) \\
        & \dot{\bm{\phi}_{\rm d}} = [\cos(\bm{\delta} ) \sodot(\bm{U}^* - \bm{K}_{\rm q} (\bm{Q} - \bm{Q}^*) + \bm{u}_{\rm q})] - \bm{v}_{\rm od}  \\
        & \dot{\bm{\phi}_{\rm q}} = [\sin(\bm{\delta} ) \sodot(\bm{U}^* - \bm{K}_{\rm q} (\bm{Q} - \bm{Q}^*) + \bm{u}_{\rm q})] - \bm{v}_{\rm oq}  \\ 
        & \dot{\bm{\gamma}_{\rm d}} = \bm{F} \sodot \bm{i}_{\rm od} \!-\! \omega_{\rm n} \bm{C}_{\rm f}  \sodot \bm{v}_{\rm oq} \!+\! \bm{K}_{\rm pv} \sodot ([\cos(\bm{\delta} ) \sodot(\bm{U}^* - \bm{K}_{\rm q} (\bm{Q} - \bm{Q}^*) + \bm{u}_{\rm q})] \!-\! \bm{v}_{\rm od}) \!+\! \bm{K}_{\rm iv} \sodot \bm{\phi}_{\rm d} - \bm{i}_{\rm ld}  \\
        & \dot{\bm{\gamma}_{\rm q}} = \bm{F} \sodot \bm{i}_{\rm oq} \!+\! \omega_{\rm n} \bm{C}_{\rm f}  \sodot \bm{v}_{\rm od} \!+\! \bm{K}_{\rm pv} \sodot ([\sin(\bm{\delta} ) \sodot(\bm{U}^* - \bm{K}_{\rm q} (\bm{Q} - \bm{Q}^*) + \bm{u}_{\rm q})] \!-\! \bm{v}_{\rm oq}) \!+\! \bm{K}_{\rm iv} \sodot \bm{\phi}_{\rm q} - \bm{i}_{\rm lq}  \\ 
        & \dot{\bm{i}_{\rm ld}} \!=\!  [ \!-\! \bm{r}_{\rm f}  \sodot \bm{i}_{\rm ld}  \!+\! \bm{K}_{\rm pc} \sodot (\bm{F} \sodot \bm{i}_{\rm od} \!-\! \omega_{\rm n} \bm{C}_{\rm f}  \sodot \bm{v}_{\rm oq} \!+\! \bm{K}_{\rm pv} \sodot ([\cos(\bm{\delta} ) \sodot(\bm{U}^* \!\!-\! \bm{K}_{\rm q} (\bm{Q} \!-\! \bm{Q}^*) \!+\! \bm{u}_{\rm q})] \!-\! \bm{v}_{\rm od}) \!+\! \bm{K}_{\rm iv} \sodot \bm{\phi}_{\rm d}  \!-\! \bm{i}_{\rm ld}) \!+\! \bm{K}_{\rm ic} \sodot \bm{\gamma}_{\rm d} \!-\! \bm{v}_{\rm od} ]\soslash \bm{L}_{\rm f}   \\
        & \dot{\bm{i}_{\rm lq}} \!=\! [\!- \!\bm{r}_{\rm f} \sodot \bm{i}_{\rm lq}  \!+\! \bm{K}_{\rm pc} \sodot (\bm{F} \sodot \bm{i}_{\rm oq} \!+\! \omega_{\rm n} \bm{C}_{\rm f}  \sodot \bm{v}_{\rm od} \!+\! \bm{K}_{\rm pv} \sodot ([\sin(\bm{\delta} ) \sodot(\bm{U}^* \!\!-\! \bm{K}_{\rm q} (\bm{Q} \!-\! \bm{Q}^*) \!+\! \bm{u}_{\rm q})] \!-\! \bm{v}_{\rm oq}) \!+\! \bm{K}_{\rm iv} \sodot \bm{\phi}_{\rm q} \!-\! \bm{i}_{\rm lq}) \!+\! \bm{K}_{\rm ic} \sodot \bm{\gamma}_{\rm q} \!-\! \bm{v}_{\rm oq} ]\soslash \bm{L}_{\rm f}  \label{eq-7-1-2:9} \\
        & \dot{\bm{v}_{\rm od}} = \omega_{\rm n} \bm{v}_{\rm oq} + \bm{i}_{\rm ld} \soslash \bm{C}_{\rm f} - \bm{i}_{\rm od} \soslash \bm{C}_{\rm f} \label{eq-7-1-2:10} \\
        & \dot{\bm{v}_{\rm oq}} = - \omega_{\rm n} \bm{v}_{\rm od} + \bm{i}_{\rm lq} \soslash \bm{C}_{\rm f} - \bm{i}_{\rm oq} \soslash \bm{C}_{\rm f} \\
        & \dot{\bm{i}_{\rm od}} = [- \bm{r}_{\rm c} \sodot \bm{i}_{\rm od} + \omega_{\rm n} \bm{i}_{\rm oq} \sodot \bm{L}_{\rm c} + \bm{v}_{\rm od} - \bm{E}_{\rm C} \bm{v}_{\rm Bd} ] \soslash \bm{L}_{\rm c} \\
        & \dot{\bm{i}_{\rm oq}} = [- \bm{r}_{\rm c} \sodot \bm{i}_{\rm oq} - \omega_{\rm n} \bm{i}_{\rm od} \sodot \bm{L}_{\rm c} + \bm{v}_{\rm oq} - \bm{E}_{\rm C} \bm{v}_{\rm Bq} ] \soslash \bm{L}_{\rm c} \label{eq-7-1-2:13}
    \end{align}
\end{subequations}
}
}

\subsection{Load model}

The loads are modelled as constant impedances, thus with the following dynamics:
\begin{subequations}\label{eq-7-1-3}
    \begin{align}
        & \dot{\bm{i}_{\rm Ld}} = - \bm{r}_{\rm L} \soslash \bm{L}_{\rm L} \sodot \bm{i}_{\rm Ld} + \omega_{\rm n} \bm{i}_{\rm Lq} + \bm{E}_{\rm L} \bm{v}_{\rm Bd} \soslash \bm{L}_{\rm L} \\
        & \dot{\bm{i}_{\rm Lq}} = - \bm{r}_{\rm L} \soslash \bm{L}_{\rm L} \sodot \bm{i}_{\rm Lq} - \omega_{\rm n} \bm{i}_{\rm Ld} + \bm{E}_{\rm L} \bm{v}_{\rm Bq} \soslash \bm{L}_{\rm L} 
    \end{align}
\end{subequations}

\subsection{Network model}

The dynamics of branches are as follows:
\begin{subequations}\label{eq-7-1-4}
    \begin{align}
        & \dot{\bm{i}_{\rm Bd}} = - \bm{r}_{\rm B} \soslash \bm{L}_{\rm B} \sodot \bm{i}_{\rm Bd} + \omega_{\rm n} \bm{i}_{\rm Bq} + \bm{E}_{\rm B} \bm{v}_{\rm Bd} \soslash \bm{L}_{\rm B} \\
        & \dot{\bm{i}_{\rm Bq}} = - \bm{r}_{\rm B} \soslash \bm{L}_{\rm B} \sodot \bm{i}_{\rm Bq} - \omega_{\rm n} \bm{i}_{\rm Bd} + \bm{E}_{\rm B} \bm{v}_{\rm Bq} \soslash \bm{L}_{\rm B}
    \end{align}
\end{subequations}

The Kirchhoff's Current Law of each bus yields
\begin{subequations}\label{eq-7-1-5}
    \begin{align}
        & \bm{0} = \bm{E}_{\rm C}\T \bm{i}_{\rm od} - \bm{E}_{\rm L}\T \bm{i}_{\rm Ld} - \bm{E}_{\rm B}\T \bm{i}_{\rm Bd} \\
        & \bm{0} = \bm{E}_{\rm C}\T \bm{i}_{\rm oq} - \bm{E}_{\rm L}\T \bm{i}_{\rm Lq} - \bm{E}_{\rm B}\T \bm{i}_{\rm Bq} 
    \end{align}
\end{subequations}

\subsection{Full model in ODE form}

Now we obtain the full dynamic model of the MG, in DAE form, as follows:
\begin{subequations}\label{eq-7-1-6}
    \begin{align}
        & \dot{\tilde{\bm{x}}} = \tilde{f}(\tilde{\bm{x}}, \bm{z} ) \label{eq-7-1-6:1}\\ 
        & \bm{0} =  \tilde{g}(\tilde{\bm{x}})  \label{eq-7-1-6:2}
    \end{align}
\end{subequations}
where (\ref{eq-7-1-6:1}) collects (\ref{eq-7-1-2}), (\ref{eq-7-1-3}), and (\ref{eq-7-1-4}); (\ref{eq-7-1-6:2}) represents (\ref{eq-7-1-5}); and 
\begin{subequations}\label{eq-7-1-7}
    \begin{align}
        & \tilde{\bm{x}} = [ \bm{\delta} ; \bm{P} ; \bm{Q} ; \bm{\phi}_{\rm d} ; \bm{\phi}_{\rm q} ; \bm{\gamma}_{\rm d} ; \bm{\gamma}_{\rm q} ; \bm{i}_{\rm ld} ; \bm{i}_{\rm lq}, \bm{v}_{\rm od} ; \bm{v}_{\rm oq} ; \bm{i}_{\rm od} ; \bm{i}_{\rm oq} ; \bm{i}_{\rm Ld} ; \bm{i}_{\rm Lq} ; \bm{i}_{\rm Bd} ; \bm{i}_{\rm Bq}] \label{eq-7-1-7:1} \\
        & \bm{z}   = [ \bm{v}_{\rm Bd} ; \bm{v}_{\rm Bq}]  \label{eq-7-1-7:2}
    \end{align}
\end{subequations}

Next, we transform the DAE model to a state-space ODE model. First, taking the derivative of both sides of (\ref{eq-7-1-7:2}) and dividing (\ref{eq-7-1-7:1}) accordingly gives
\begin{subequations}\label{eq-7-1-8}
    \begin{align}
        & \dot{\tilde{\bm{x}}}_1 = f_1(\tilde{\bm{x}}_1, \tilde{\bm{x}}_2, \bm{z} )  \label{eq-7-1-8:1} \\
        & \dot{\tilde{\bm{x}}}_2 = f_2(\tilde{\bm{x}}_1, \tilde{\bm{x}}_2, \bm{z}) \label{eq-7-1-8:2} \\
        & \bm{0} = \tilde{g}_{\rm der} (\tilde{\bm{x}}_2, \bm{z}) =  \bm{M}_1 \tilde{\bm{x}}_2 + \bm{M}_2 \bm{z}  \label{eq-7-1-8:3}
    \end{align}
\end{subequations}
where (\ref{eq-7-1-8:1}) collects (\ref{eq-7-1-2:1})-(\ref{eq-7-1-2:9}); (\ref{eq-7-1-8:2}) collects (\ref{eq-7-1-2:10})-(\ref{eq-7-1-2:13}), (\ref{eq-7-1-3}), and (\ref{eq-7-1-4}); and 
\begin{subequations}\label{eq-7-1-9}
    \begin{align}
        & \tilde{\bm{x}}_1 = [ \bm{\delta} ; \bm{P} ; \bm{Q} ; \bm{\phi}_{\rm d} ; \bm{\phi}_{\rm q} ; \bm{\gamma}_{\rm d} ; \bm{\gamma}_{\rm q} ; \bm{i}_{\rm ld} ; \bm{i}_{\rm lq} ]  \\
        & \tilde{\bm{x}}_2 = [\bm{v}_{\rm od} ; \bm{v}_{\rm oq} ; \bm{i}_{\rm od} ; \bm{i}_{\rm oq} ; \bm{i}_{\rm Ld} ; \bm{i}_{\rm Lq} ; \bm{i}_{\rm Bd} ; \bm{i}_{\rm Bq}]   
    \end{align}  
\end{subequations}

Based on (\ref{eq-7-1-8:3}), we have 
\begin{equation}
    \bm{z}  = - \bm{M}_2^{-1} \bm{M}_1 \tilde{\bm{x}}_2
\end{equation}
Substituting this equation into (\ref{eq-7-1-6:1}) gives the ODE model as follows:
\begin{equation}
    \dot{\tilde{\bm{x}}} = \tilde{f}(\tilde{\bm{x}},  - \bm{M}_2^{-1} \bm{M}_1 \tilde{\bm{x}}_2 )
\end{equation}

Finally, we reformulate the ODE model into an affine form in terms of the renewable uncertainty and control, which gives 
\begin{equation}
    \dot{\bm{x}} = f(\bm{x}) + \bm{K} \bm{\xi} + g(\bm{x}) \bm{u}
\end{equation}
where $\bm{x} = \tilde{\bm{x}}$ is the state variable vector of the ODE model, $f$ and $g$ are functions determined by the MG dynamics, and $\bm{K}$ being a constant matrix formed by $K_{{\rm p}, i}$ and zeros. 


\section{Parameters of the MG}

The parameters of the MG are divided into two categories: time-invariant parameters and time-varying parameters. 
Time-invariant parameters mainly include the physical parameters of components and partial parameters of controllers. 
Time-varying parameters include load power, actual and forecast power outputs of generators, and the setpoints of converters. 

\subsection{Time-invariant parameters}

The values of all time-invariant parameters are listed in Table \ref{tab-7-1-ap-1} and Table \ref{tab-7-1-ap-2}. 

\begin{table}[h]
    \centering
    \setlength{\tabcolsep}{3.3pt}  
    \setlength\extrarowheight{2pt}

    \caption{Time-invariant parameters of generators}
    \begin{tabular*}{0.90\hsize}{|ccccccccccc|}\hline\hline
    Generator &   
    $\bm{r}_{\rm f} (\Omega)$   &   $\bm{L}_{\rm f} (\rm{H})$  & $\bm{C}_{\rm f} (\rm{F})$   &  $\bm{r}_{\rm c} (\Omega)$   &  $\bm{L}_{\rm c} (\rm{H})$ &  $\omega_{\rm c}$   & $\bm{K}_{\rm p}$   & $\bm{K}_{\rm q}$  & $\omega_{\rm n} (\text{rad/s})$  &  $\omega^* (\text{rad/s})$ \\ \hline 
    G1 & 0.1 &  1.35$\!\times\! 10^{-3}$ & 5$\!\times\! 10^{-5}$  &  0.1  &     1$\!\times\! 10^{-3}$  &  15.705 & 0.333$\!\times\! 10^{-6}$  & 0.333$\!\times\! 10^{-5}$  &  $100\pi$  & $100\pi$\\ 
    G2 & 0.1 &  1.35$\!\times\! 10^{-3}$ & 5$\!\times\! 10^{-5}$  &  0.1  &     1$\!\times\! 10^{-3}$  &  15.705 & 1.0  $\!\times\! 10^{-6}$  & 1.0  $\!\times\! 10^{-5}$   &  $100\pi$ & $100\pi$ \\
    G3 & 0.1 &  1.35$\!\times\! 10^{-3}$ & 5$\!\times\! 10^{-5}$  &  0.1  &     1$\!\times\! 10^{-3}$  &  15.705 & 1.25 $\!\times\! 10^{-6}$  & 1.25 $\!\times\! 10^{-5}$   &  $100\pi$ & $100\pi$ \\
    G4 & 0.1 &  1.35$\!\times\! 10^{-3}$ & 5$\!\times\! 10^{-5}$  &  0.1  &     1$\!\times\! 10^{-3}$  &  15.705 & 1.25 $\!\times\! 10^{-6}$  & 1.25 $\!\times\! 10^{-5}$   &  $100\pi$ & $100\pi$ \\
    G5 & 0.1 &  1.35$\!\times\! 10^{-3}$ & 5$\!\times\! 10^{-5}$  &  0.1  &     1$\!\times\! 10^{-3}$  &  15.705 & 1.0  $\!\times\! 10^{-6}$  & 1.0  $\!\times\! 10^{-5}$   &  $100\pi$ & $100\pi$ \\
    G6 & 0.1 &  1.35$\!\times\! 10^{-3}$ & 5$\!\times\! 10^{-5}$  &  0.1  &     1$\!\times\! 10^{-3}$  &  15.705 & 0.833$\!\times\! 10^{-6}$  & 0.833$\!\times\! 10^{-5}$   &  $100\pi$ & $100\pi$ \\
    G7 & 0.1 &  1.35$\!\times\! 10^{-3}$ & 5$\!\times\! 10^{-5}$  &  0.1  &     1$\!\times\! 10^{-3}$  &  15.705 & 0.833$\!\times\! 10^{-6}$  & 0.833$\!\times\! 10^{-5}$   &  $100\pi$ & $100\pi$ \\
    G8 & 0.1 &  1.35$\!\times\! 10^{-3}$ & 5$\!\times\! 10^{-5}$  &  0.1  &     1$\!\times\! 10^{-3}$  &  15.705 & 0.833$\!\times\! 10^{-6}$  & 0.833$\!\times\! 10^{-5}$   &  $100\pi$ & $100\pi$ \\
    G9 & 0.1 &  1.35$\!\times\! 10^{-3}$ & 5$\!\times\! 10^{-5}$  &  0.1  &     1$\!\times\! 10^{-3}$  &  15.705 & 0.667$\!\times\! 10^{-6}$  & 0.667$\!\times\! 10^{-5}$   &  $100\pi$ & $100\pi$ \\ 
    \end{tabular*} \hspace*{-0.1pt}
    \begin{tabular*}{0.90\hsize}{|cccccccccc|c} \hline\hline
        Generator &  
         $\bm{K}_{\rm pv}$  &  $\bm{K}_{\rm iv}$  &  $~\bm{K}_{\rm pc}~$  &  $~\bm{K}_{\rm ic}~$  &  $~\bm{F}~$    &  $~\bm{P}_{\rm max} (\rm{MW})~$ &  ~$\bm{P}_{\rm min} (\rm{MW})$~ & $\bm{Q}_{\rm max} (\rm{MVAR})$ & $\bm{Q}_{\rm min} (\rm{MVAR})$ $\!\!\!\!\!\!~~~~$ &  \\ \hline 
        G1 &  0.05 & 390 & 10.5 & $16 \times 10^3$ & 0.75 & 3   & -3   & 3 & -3 &   \\
        G2 &  0.05 & 390 & 10.5 & $16 \times 10^3$ & 0.75 & 1.2 & 0    & 2 & -2 &   \\
        G3 &  0.05 & 390 & 10.5 & $16 \times 10^3$ & 0.75 & 1.2 & 0    & 2 & -2 &   \\
        G4 &  0.05 & 390 & 10.5 & $16 \times 10^3$ & 0.75 & 1.2 & 0    & 2 & -2 &   \\
        G5 &  0.05 & 390 & 10.5 & $16 \times 10^3$ & 0.75 & 1.5 & -1.5 & 2 & -2 &   \\
        G6 &  0.05 & 390 & 10.5 & $16 \times 10^3$ & 0.75 & 1   & -1   & 1 & -1 &   \\
        G7 &  0.05 & 390 & 10.5 & $16 \times 10^3$ & 0.75 & 0.8 & 0    & 1 & -1 &   \\
        G8 &  0.05 & 390 & 10.5 & $16 \times 10^3$ & 0.75 & 0.8 & 0    & 1 & -1 &   \\
        G9 &  0.05 & 390 & 10.5 & $16 \times 10^3$ & 0.75 & 1   & -1   & 1 & -1 &   \\  \hline\hline
        \end{tabular*}
    \label{tab-7-1-ap-1}  
\end{table}

\begin{table}[h]
    \centering
    \setlength{\tabcolsep}{3pt}  
    \setlength\extrarowheight{2pt}

    \caption{Parameters of lines}
    \begin{tabular}{|cccc|cccc|}\hline\hline
    From bus & To bus & $\bm{r}_{\rm B} (\Omega)$ & $\bm{L}_{\rm B} (\rm{H})$  & From bus & To bus & $\bm{r}_{\rm B} (\Omega)$ & $\bm{L}_{\rm B} (\rm{H})$   \\ \hline 
    2           & 3   & 0.493   &  0.000799   &   2	          & 19  & 0.164   &  0.000498     \\      
    3           & 4   & 0.366   &  0.000593   &   19          & 20  & 1.5042  &  0.004314       \\       
    4           & 5   & 0.3811  &  0.000618   &   20          & 21  & 0.4095  &  0.001523       \\       
    5           & 6   & 0.819   &  0.00225    &   21          & 22  & 0.7089  &  0.002984       \\       
    6           & 7   & 0.1872  &  0.00197    &   3	          & 23  & 0.4512  &  0.000981     \\       
    7           & 8   & 0.7114  &  0.000748   &   23          & 24  & 0.898   &  0.002257       \\       
    8           & 9   & 1.03    &  0.002355   &   24          & 25  & 0.896   &  0.002232       \\       
    9           & 10  & 1.044   &  0.002355   &   6	          & 26  & 0.203   &  0.000329     \\        
    10          & 11  & 0.1966  &  0.000207   &   26          & 27  & 0.2842  &  0.000461       \\        
    11          & 12  & 0.3744  &  0.000394   &   27          & 28  & 1.059   &  0.002972       \\        
    12          & 13  & 1.468   &  0.003676   &   28          & 29  & 0.8042  &  0.00223        \\        
    13	        & 14  & 0.5416  &  0.002269   &   29          & 30  & 0.5075  &  0.000823       \\        
    14	        & 15  & 0.591   &  0.001674   &   30          & 31  & 0.9744  &  0.003065       \\        
    15          & 16  & 0.7463  &  0.001735   &   31          & 32  & 0.3105  &  0.001152       \\        
    16   	    & 17  & 1.289   &  0.005478   &   32          & 33  & 0.341   &  0.001688       \\        
    17	        & 18  & 0.732   &  0.001827   &  & & &  \\  \hline\hline      
    \end{tabular} 
    \label{tab-7-1-ap-2}  
\end{table}

\subsection{Time-varying parameters}

\begin{figure}[b]
	\centering
 	\includegraphics[scale=0.45]{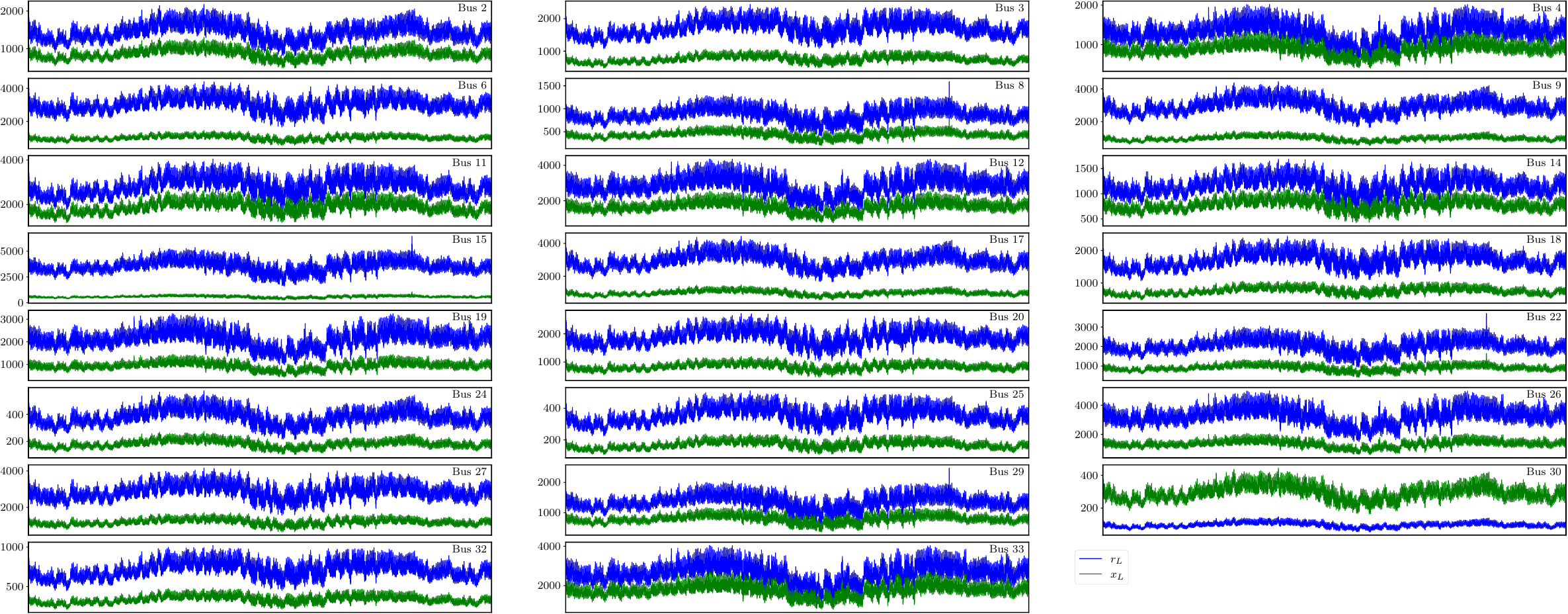}  
	\caption{Profiles of load impedance of one year.}
	\label{fig-7-1-ap-r1-full}
\end{figure}

\begin{figure}[h]
	\centering
 	\includegraphics[scale=0.45]{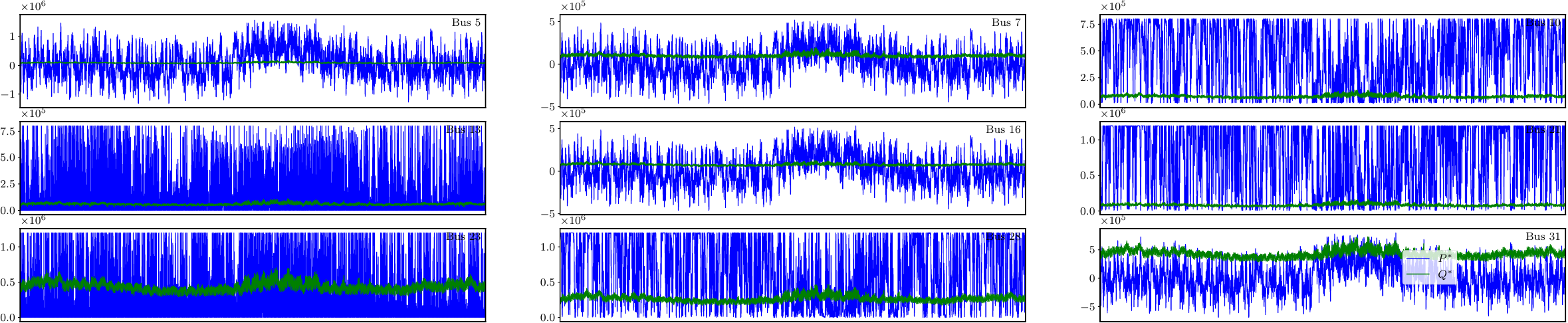}  
	\caption{Profiles of power setpoints of all converters of one year.}
	\label{fig-7-1-ap-r2-full}
\end{figure}

\begin{figure}[h]
	\centering
 	\includegraphics[scale=0.45]{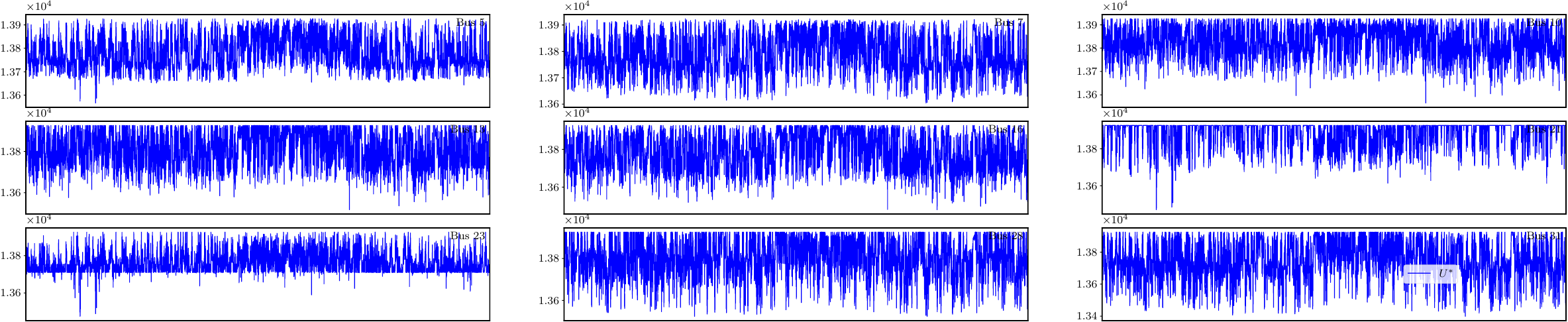}  
	\caption{Profiles of voltage magnitude setpoints of all converters of one year.}
	\label{fig-7-1-ap-r2v-full}
\end{figure}

\begin{figure}[h]
	\centering
 	\includegraphics[scale=0.45]{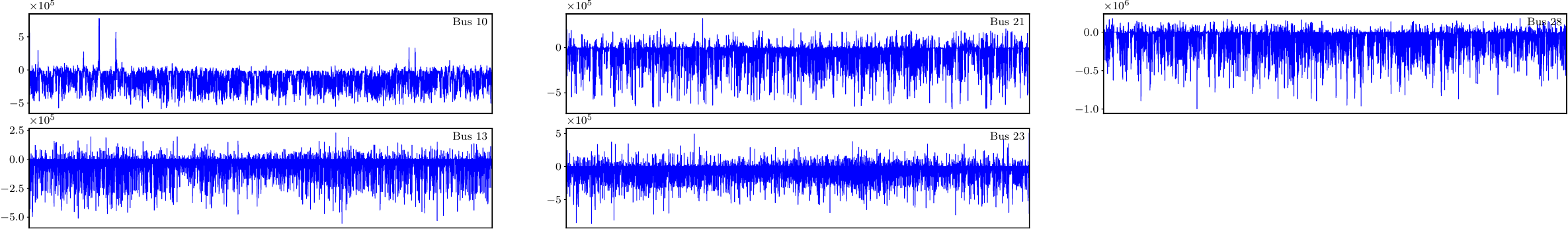}  
	\caption{Profiles of forecast errors of wind and solar panel generators of one year.}
	\label{fig-7-1-ap-r3-full}
\end{figure}

\begin{figure}[h]
	\centering
 	\includegraphics[scale=0.5]{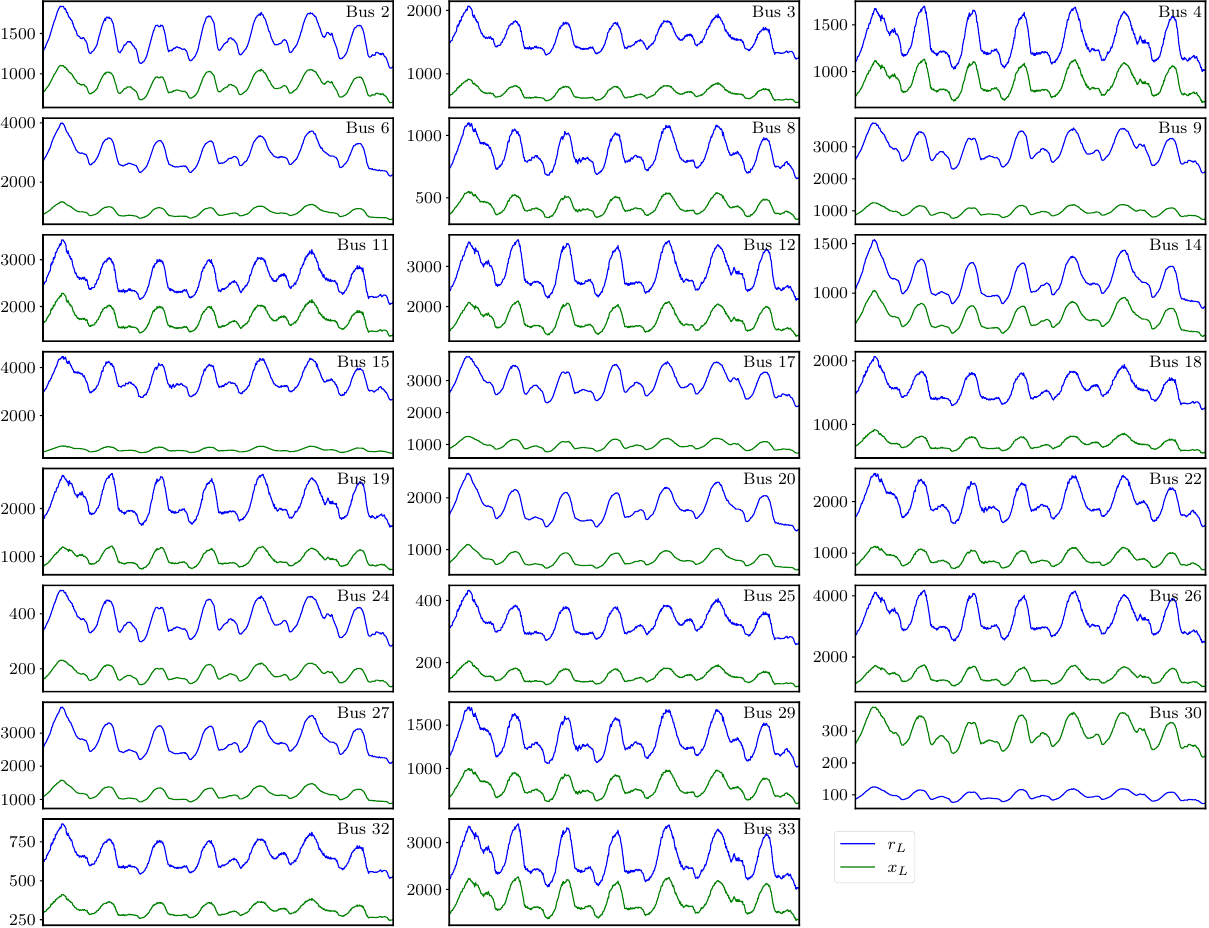}  
	\caption{Profiles of load impedance of one week.}
	\label{fig-7-1-ap-r1}
\end{figure}

\begin{figure}[h]
	\centering
 	\includegraphics[scale=0.5]{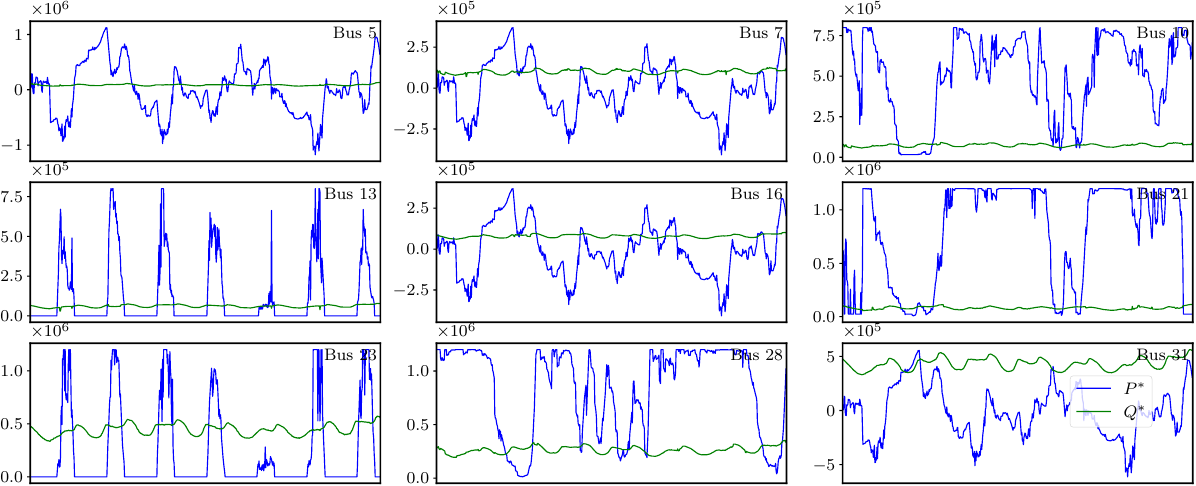}  
	\caption{Profiles of power setpoints of all converters of one week.}
	\label{fig-7-1-ap-r2}
\end{figure}

\begin{figure}[h]
	\centering
 	\includegraphics[scale=0.5]{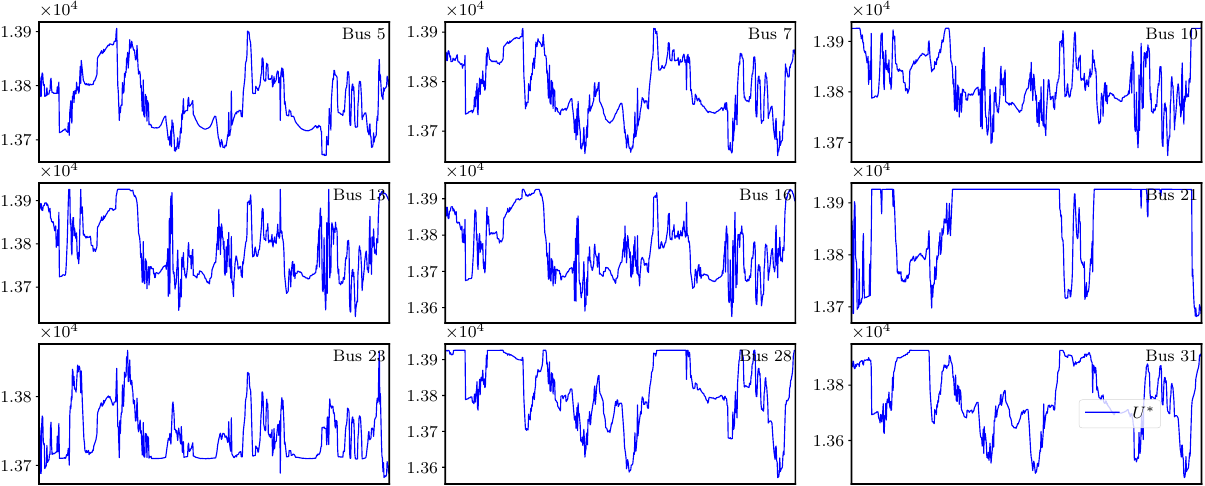}  
	\caption{Profiles of voltage magnitude setpoints of all converters of one week.}
	\label{fig-7-1-ap-r2v}
\end{figure}

\begin{figure}[h]
	\centering
 	\includegraphics[scale=0.5]{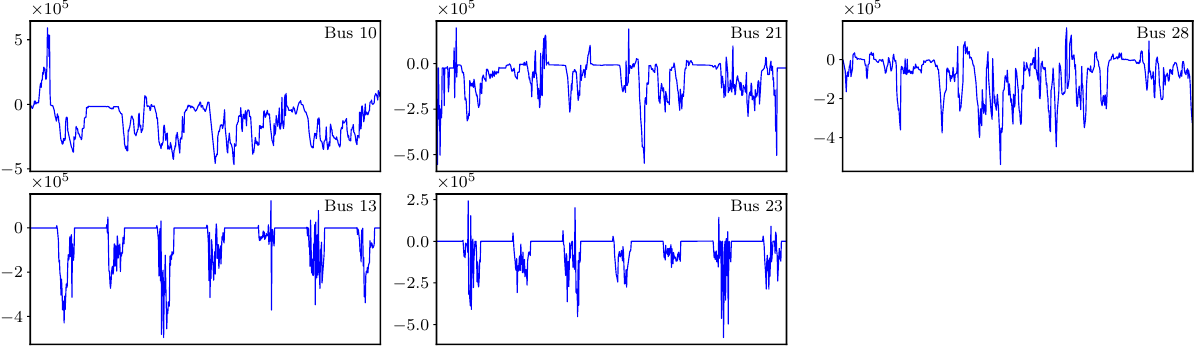}  
	\caption{Profiles of forecast errors of wind and solar panel generators of one week.}
	\label{fig-7-1-ap-r3}
\end{figure}

Time-varying parameter values of one year at 15-min resolution are generated based on the time-coincident load, wind, and solar data in the ARPA-E PERFORM datasets. The load uncertainties are omitted in our model, and thus time-varying values of $\bm{r}_{\rm L}$ and $\bm{L}_{\rm L}$ are generated using the actual load data. 
ARPA-E PERFORM datasets contain forecast and actual data of wind and solar sites, where the forecast data is used to generate time-varying setpoints $\bm{P}^*$ of wind and solar panel generators, and the deviations between the actual and forecast values are used to generated time-varying errors $\bm{\xi}$ of wind and solar panel generators. 
For energy storage systems, their setpoints $\bm{P}^*$ are set by allocating the unbalanced power (total active load power $-$ total forecast power outputs of wind and solar panel generators) in proportion to the maximal active power outputs. The setpoints $\bm{U}^*$ and $\bm{Q}^*$ are obtained by Volt/VAR optimization with the objective of minimizing power losses. The profiles of all time-varying parameters of one year are shown in Fig. \ref{fig-7-1-ap-r1-full} to Fig. \ref{fig-7-1-ap-r3-full}, and Fig. \ref{fig-7-1-ap-r1} to Fig. \ref{fig-7-1-ap-r3} show the profiles for one week.
The entire one-year data can be found in \url{https://github.com/thanever/SOC/tree/master/Data/33-bus-MG}

\vspace*{200pt}

\ifCLASSOPTIONcaptionsoff
  \newpage
\fi

\bibliographystyle{IEEEtran}
\bibliography{4}

\end{document}

%% file: Preamble/preamble.tex
\usepackage{booktabs} 
\usepackage{multirow}
\usepackage{mathtools}

\usepackage{circuitikz}

\usepackage{amssymb}
 
\usepackage{enumitem}

\usepackage{smartdiagram}
\usesmartdiagramlibrary{additions}

\graphicspath{{Figs/}}

\usepackage{pdfpages}
\usepackage{pdflscape}
\usepackage{amsmath}
\usepackage{amsthm}

\usepackage{amsfonts}

\usepackage{color}

\usepackage{setspace}
\usepackage{enumitem}
\usepackage{xcolor}

\usepackage{bm}
\usepackage{bbm}

\usepackage{tikz}
\usepackage{pgf}
\usetikzlibrary{arrows,automata}
\usetikzlibrary{shapes}
\tikzstyle{data44}=[rectangle split,rectangle split parts=2,draw,text centered]

\usepackage{subfigure}

\DeclareMathAlphabet{\mathcal}{OMS}{cmsy}{m}{n}

\usetikzlibrary{matrix}

\tikzset{
  BarreStyle/.style =   {opacity=.3,line width=14 mm,color=#1},
  node style ge/.style={},
  node style sp/.style={},
  yl/.style={},
  arrow style mul/.style={},
}

\newtheoremstyle{mytheoremstyle} 
        {\topsep}                    
        {\topsep}                    
        {\fontfamily{ptm}\selectfont}                   
        {}                           
        {\itshape\fontfamily{ptm}\selectfont}                   
        {:}                          
        {.5em}                       
        {}  
\theoremstyle{mytheoremstyle}

\usepackage{threeparttable,booktabs}

\usepackage[colorlinks,
            linkcolor=blue,
            anchorcolor=blue,
            citecolor=blue
            ]{hyperref}

\usepackage{cite}

\usepackage{mdwlist}

\usepackage[linesnumbered,ruled,vlined]{algorithm2e}
\SetKwInput{KwInput}{Input}                
\SetKwInput{KwOutput}{Output}   
\let\oldnl\nl
\newcommand{\nonl}{\renewcommand{\nl}{\let\nl\oldnl}}

\usepackage{etoolbox}

\makeatletter
\patchcmd{\@algocf@start}
  {-1.5em}
  {0pt}
  {}{}
\makeatother

\usepackage{textcomp}

\usepackage{stmaryrd}

\usepackage{comment}

\usepackage{graphicx,subfigure}

\usepackage{stfloats}
\usepackage{tabularx }

\newcommand\smallmath[2]{#1{\raisebox{\dimexpr \fontdimen 22 \textfont 2
      - \fontdimen 22 \scriptscriptfont 2 \relax}{$\scriptscriptstyle #2$}}} 
\newcommand\sodot{\smallmath\mathbin{\!\!\odot\!\!}}
\newcommand\soslash{\smallmath\mathbin{\!\!\oslash\!\!}}

\usepackage{makecell}

\usepackage{colortbl}

\newcolumntype{P}[1]{>{\centering\arraybackslash}p{0.66cm}}
 
\usepackage{anyfontsize}
\usepackage{t1enc}

\newcommand{\T}{^{\scriptscriptstyle\rm T}}

\usepackage[strict]{changepage}
\usepackage{relsize}
 
\usepackage{nccmath}

\usepackage{arydshln}